\documentclass[aps,pre,reprint,superscriptaddress,floatfix,longbibliography]{revtex4-1}

\usepackage[utf8]{inputenc}
\usepackage{amsmath,amsfonts,amssymb}
\usepackage{bm}
\usepackage{siunitx}
\usepackage{graphicx}
\usepackage{stmaryrd}
\usepackage{xcolor}
\usepackage[normalem]{ulem}
\usepackage{cancel}
\usepackage[colorlinks=true,citecolor=blue,urlcolor=blue]{hyperref}
\usepackage{etoolbox}   
\usepackage{orcidlink}

\newcommand{\Mean}[1]{\left\langle #1 \right\rangle}
\newcommand{\mean}[1]{\langle #1 \rangle}
\newcommand{\eff}{\mathrm{eff}}
\newcommand{\Q}{\mathcal{Q}}

\newcommand{\kt}{\kB T}
\newcommand{\kB}{k_{\rm B}}
\newcommand{\tf}{t_{\rm f}}
\newcommand{\tr}{t_{\rm r}}
\newcommand{\dd}{\mathrm{d}}

\newcommand{\PW}{\mathcal{P}}

\DeclareMathOperator{\e}{e}

\newcommand{\FOUNDRYLBL}{%
	Molecular Foundry, Lawrence Berkeley National Laboratory, Berkeley, CA 94720, USA}

\newcommand{\LPENSL}{%
	\href{https://ror.org/02feahw73}{CNRS},
	\href{https://ror.org/04zmssz18}{ENS de Lyon},
	\href{https://ror.org/00w5ay796}{Laboratoire de Physique},
	F-69342 Lyon, France}

\DeclareRobustCommand{\demonmark}{%
	\raisebox{-0.25ex}{\includegraphics[height=2.5ex]{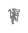}}}

\DeclareRobustCommand{\lyonmark}{%
	\raisebox{-0.25ex}{\includegraphics[height=2.5ex]{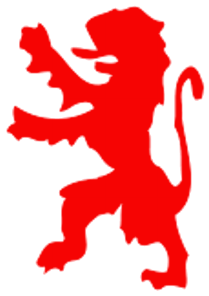}}}

\makeatletter
\def\@fnsymbol#1{%
	\ifcase#1\or
	\mbox{\demonmark}\or
	\mbox{\lyonmark}\or
	\ddagger\or
	\S\or
	\P\or
	\|\or
	**\or
	\dagger\dagger
	\else
	\@ctrerr
	\fi}
\makeatother

\def\c#1{~\cite{#1}}

\def\cc#1{~Ref.~\onlinecite{#1}}
\def\ccc#1{~Refs.~\onlinecite{#1}}

\def\av#1{\langle #1 \rangle}

\def\beq{\begin{equation}}
	\def\eeq{\end{equation}}
\def\bea{\begin{eqnarray}}
	\def\eea{\end{eqnarray}}

\def\eqq#1{Eq.~(\ref{#1})}

\def\eq#1{(\ref{#1})}

\def\f#1{Fig.~\ref{#1}}

\def\s#1{Section~\ref{#1}}
\def\a#1{Appendix~\ref{#1}}

\begin{document}

\title{A neural-network Maxwell's demon learns cold damping for work extraction}

\author{Stephen Whitelam\,\orcidlink{0000-0002-0086-6803}}
\email{swhitelam@lbl.gov}
\affiliation{\FOUNDRYLBL}
\author{Sergio Ciliberto\, \orcidlink{0000-0002-4366-6094}}
\affiliation{\LPENSL}
\author{Ludovic Bellon\, \orcidlink{0000-0002-2499-8106}}
\email{ludovic.bellon@ens-lyon.fr}
\affiliation{\LPENSL}

\date{\today}

\begin{abstract}

We train a neural-network Maxwell's demon to extract work from a model of an underdamped micromechanical cantilever subject to thermal noise. The demon, which periodically adjusts the position of a harmonic trap, is trained to maximize the power extracted under steady-state operation. When the demon is given the cantilever position and trap position as inputs it learns a refined version of an existing hand-designed protocol, yielding a substantial improvement in performance. When the demon receives the oscillator velocity as input it discovers a qualitatively different strategy that extracts substantially more work, close to the theoretical power bound. Analysis of the protocol shows that it implements {\em cold damping}: the trap position is displaced approximately linearly with velocity, producing an effective increase of the oscillator's damping coefficient and a reduction of its effective temperature. Thus a neural-network Maxwell's demon rediscovers a well-known cooling strategy from optomechanics, revealing a simple physical mechanism underlying near-optimal work extraction from thermal fluctuations in an underdamped system.
\end{abstract}

\maketitle

\section{Introduction}

Feedback control can be used to extract energy from thermal fluctuations. By measuring a fluctuating degree of freedom and adjusting a control parameter in response, we can rectify thermal motion and extract work from a single heat bath, a process embodied by Maxwell's demon\c{maxwelltheory} and the Szilard engine\c{szilard1964decrease}. Such information engines\c{Archambault-2025, saha2021maximizing} respect a generalized second law and associated fluctuation theorems\c{Sagawa_2010,Parrondo_2015}, and they have been realized in a range of laboratory experiments\c{Toyabe_2010}.

An important open question is how to control such systems optimally: which feedback rule extracts the most work, or dissipates the least, given the constraints at hand? Here we study a paradigmatic example in which a protocol discovered by evolutionary learning turns out to have a simple and illuminating physical interpretation. This simplicity is a surprise, given that optimal protocols in general can be difficult to interpret physically, often containing discontinuous jumps and other nonanalytic features\c{Seifert_2012,blaber2021steps}.

We consider an underdamped oscillator in a harmonic trap, a model of a micromechanical cantilever\c{Dago-2022-JStat, Archambault-2024, Barros-2025}, coupled to a thermal bath. At regular intervals a feedback rule updates the trap position; we refer to this rule as a {\em demon}, in the sense of Maxwell. We encode the demon as a deep neural network and train it, by genetic algorithm, to extract as much power as possible in steady-state operation\c{whitelam2023demon, Barros-2025}.

Given access to the oscillator and trap positions the demon learns a refined version of an existing hand-designed protocol\c{Archambault-2024}, extracting about 50\% more power. Given instead the oscillator velocity it discovers a qualitatively different strategy that extracts substantially more work, almost saturating the thermodynamic bound on power extraction. We show that this strategy implements {\em cold damping}: the trap is displaced approximately linearly with velocity, which raises the oscillator's effective damping coefficient and lowers its effective temperature. In maximizing work extraction the network thus rediscovers a cooling technique well known from optomechanics\c{Mancini-1998, Poggio-2007, Aspelmeyer-2014}. We use the same description to predict the full distribution of extracted work, including its characteristic skew, and to compare the energetics and entropy production of work-extracting velocity feedback with those of feedback cooling\c{Munakata_2012,Rosinberg_PRE_2017, Kim_PRL_2004, Kim_PRE_2007}. We find that the mean extracted power does not depend on whether the feedback force is treated as internal or external to the system, whereas its fluctuations do. We also verify that the relevant integral fluctuation theorems for heat and entropy production continue to hold at the finite feedback intervals used in simulation.

The effectiveness of velocity feedback emphasizes the role of inertia in underdamped systems\c{Archambault-2024}. Inertial effects are also central to {\em momentum computing}, in which information is stored and processed in the velocity degree of freedom\c{ray2021non,ray2023gigahertz}. More broadly, our results illustrate how machine learning can do more than optimize a control protocol: by admitting a simple physical interpretation, the learned solution reveals the mechanism responsible for near-optimal operation, and suggests the possibility of interpretable control strategies in other fluctuating nanoscale systems.

\section{Model and work-extraction protocols} \label{Sec_model}

Our simulation model of the micromechanical cantilever of\ccc{Dago-2022-JStat, Dago-2024-Chapter, Barros-2025} consists of an oscillator, specified by a position $x$, in a harmonic trap of spring constant $k$ and center $x_0$. This system is sketched in \f{fig1}(a). The oscillator evolves in time according to the underdamped Langevin equation 
\beq
\label{lang}
m \ddot x + \gamma \dot x +k(x-x_0) = \sqrt{2\gamma \kt} \, \xi(t),
\eeq
where $m$ is the oscillator mass, $\gamma$ is the damping coefficient, and $\xi(t)$ is a Gaussian white noise with correlations $\av{\xi(t)}=0$ and $\av{\xi(t) \xi(t')}=\delta(t-t')$. We choose parameters appropriate to the experiments of\ccc{Archambault-2024, Archambault-2025, Barros-2025}: the basic length scale is $\sigma = \sqrt{\kt/k} \approx \SI{1}{nm}$, the basic time scale is $\omega_0^{-1} = \sqrt{m/k} \approx 0.14\, {\rm ms}$, and the basic energy scale is $\kt$ (where $k_{\rm B}$ is the Boltzmann constant and $T= \SI{295}{K}$). The quality factor of the oscillator is $Q_f =m \omega_0/\gamma =10$, and its energy relaxation time is $\tr = Q_f \omega_0^{-1} \approx 1.4 \, {\rm ms}$. We simulate \eqq{lang} using the integration scheme of\cc{Barros-2025}.

\begin{figure}[t]
\begin{center}
\includegraphics[width=\linewidth]{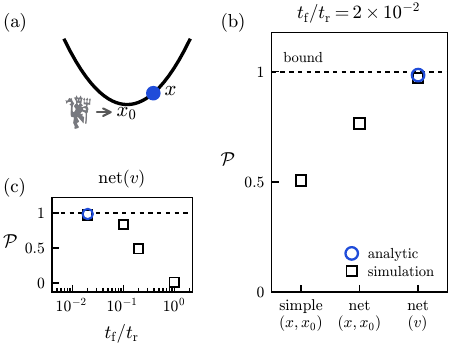}
\end{center}
\caption{(a) Schematic of an underdamped oscillator with position $x$ in a harmonic trap centered at $x_0$. The trap position $x_0$ is updated periodically by a Maxwell's demon that enacts one of three feedback-control protocols, defined by Eqs.~\eq{simple},~\eq{netx}, or~\eq{netv}. (b) Average power $\PW$ (in units of $\kt/\tr$) extracted by the three protocols (squares). The neural-network protocol that takes the oscillator velocity $v$ as input almost saturates the bound given by \eqq{meanQdot} (dotted line). This protocol enacts {\em cold damping}, leading to an analytic estimate for power extraction for continuous feedback given by \eqq{analytic} (circle). (c) Power as a function of feedback time for the neural-network $v$-protocol (each protocol is trained at the indicated feedback times).}
\label{fig1}
\end{figure}

Starting in thermal equilibrium with an initial trap position $x_0(0)$, we apply a feedback control protocol to $x_0$ in order to extract work from the system. At regular time intervals, spaced by $t_{\rm f}=2\times10^{-2} \tr$, the position $x_0$ of the trap is updated by one of three protocols. The system quickly reaches a nonequilibrium steady state, which we analyze to characterize the power extraction by the demon.

In the first protocol, the trap position is initialized to $x_0(0) = L$. Thereafter, at regular intervals, the protocol takes the current trap position $x_0^-$ and the cantilever position $x$, and sets the new trap position $x_0^+$ to
\beq
\label{simple}
x_0^+ =
\begin{cases}
L & \text{if} \ x_0^- =-L \ \text{and}\ x > h, \\
-L & \text{if} \ x_0^- =L \ \text{and}\ x < -h, \\
x_0^- & \text{otherwise}.
\end{cases}
\eeq

Thus the trap position flips sign when the oscillator crosses a threshold on the opposite side of the origin to the current trap position. The two parameters of this protocol are the location parameter $L$ and the threshold parameter $h$. 
\begin{figure*}[t]
\begin{center}
\includegraphics[width=\linewidth]{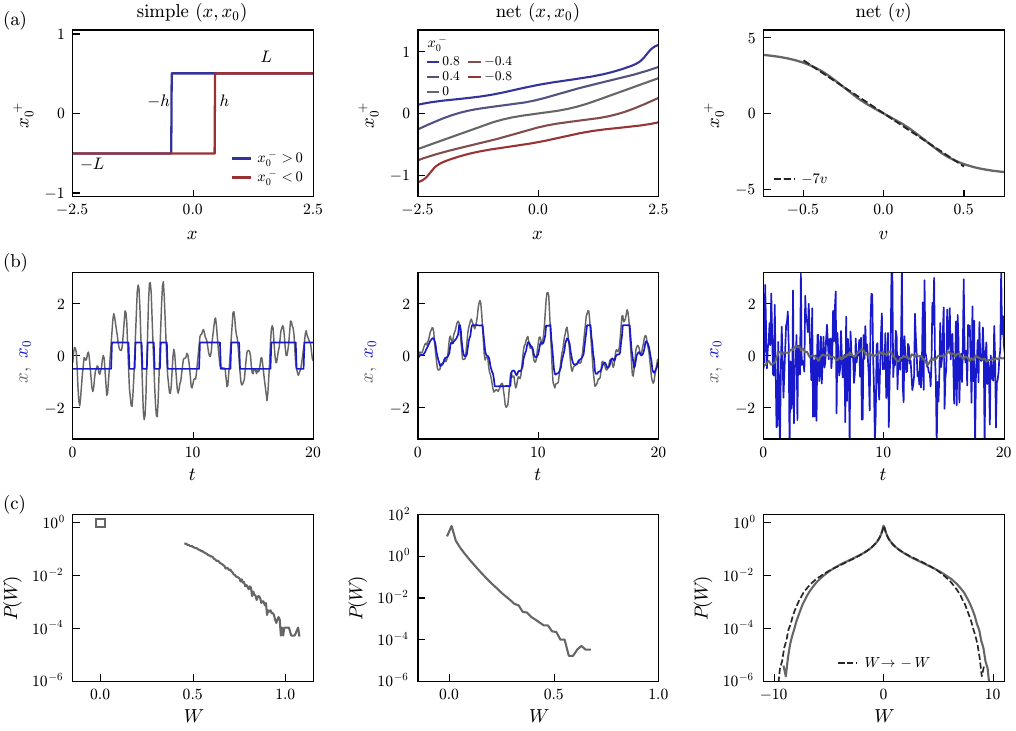}
\end{center}
\caption{For the three feedback-control protocols, \eqq{simple} (left panels), \eqq{netx} (center panels), and \eqq{netv} (right panels), we show (a) the protocol, i.e. the new value of the trap position $x_0$ as a function of the protocol inputs; (b) time traces of oscillator position $x$ and trap position $x_0$; and (c) probability distributions of work $W$ for each feedback event. In (c, right) the black dashed line shows the reflection in the vertical axis of the probability distribution, revealing its slight asymmetry. In this figure $x$ and $x_0$ are shown in units of $\sigma$, $v$ is shown in units of $\sigma \omega_0$, $t$ is shown in units of $\omega_0^{-1}$, and $W$ is shown in units of $\kt$.}
\label{fig2}
\end{figure*}

The second protocol also takes as input the current trap position $x_0^-$ and oscillator position $x$. The trap position is initialized to $x_0(0)=0$, and every $\tf$ the new trap position is set to
\beq
\label{netx}
x_0^+ = f_{\bm \theta}(x,x_0^-),
\eeq
where $f$ is the output of a deep neural network whose parameters (weights and biases) are ${\bm \theta}$. 

The third protocol, which also begins from $x_0(0)=0$, takes instead the oscillator velocity $v=\dot{x}$ as input. Every $\tf$ the new trap position is set to
\beq
\label{netv}
x_0^+ = g_{\bm \theta}(v),
\eeq
where $g$ is the output of a deep neural network whose parameters are ${\bm \theta}$. 

We simulate trajectories of duration $\tau$. The average power extracted~\footnote{In this article, we use the engine convention for work and heat: work extracted from the system  and heat coming from the bath are positive quantities.} by the feedback is
\beq
\label{pow}
\PW= \frac{1}{\tau}
\sum_{n=1}^{N} W_n, \quad W_n=U(x,x_{0}^-)-U(x,x_{0}^+),
\eeq
where the work $W_n$ is the change of potential energy
\beq
U(x,x_0)=\frac{1}{2}k(x-x_0)^2
\eeq
due to the $n^{\rm th}$ feedback event. The trajectory duration is $\tau = 6\times10^{4}\,\tr$. Given that feedback is applied at time intervals $t_{\rm f}=2\times 10^{-2}\, \tr$, the number of feedback events is $N= 3\times10^{6}$. 

We trained each protocol by genetic algorithm in order to maximize $\PW$, using the methods described in detail in\ccc{Barros-2025,whitelam2023demon,whitelam2025benchmark}. For the simple protocol this amounts to a search over the two parameters $h$ and $L$. For the neural-network protocols this amounts to a search over the parameters ${\bm \theta}$ that determine the functions $f$ and $g$.

We first considered unrestricted functions $f$ and $g$, and found that the learned protocols approximately obeyed the symmetries $f_{\bm \theta}(-x,-x_0) = -f_{\bm \theta}(x,x_0)$ and $g_{\bm \theta}(-v) = -g_{\bm \theta}(v)$. These symmetries reflect the fact that the oscillator dynamics is invariant under the transformation
\beq
(x,v,x_0) \leftrightarrow (-x,-v,-x_0),
\eeq
which corresponds to reflecting the coordinate system about the origin.

We then imposed these symmetries on $f$ and $g$ to determine whether the maximum extracted power $\PW$ was affected. We found that it was not, although learning with the constrained functions was faster than with the unconstrained ones. For simplicity we present in the following section the results obtained using the constrained functions.

\section{Trained protocols}

In \f{fig1}(b) we show the average power extracted under the three protocols \eq{simple}, \eq{netx}, and \eq{netv}, following genetic optimization of their parameters. The simple protocol \eq{simple} takes the oscillator position $x$ and the current trap position $x_0$ as input, and is optimized by the parameter choices $h \approx 0.451\sigma$ and $L \approx 0.504 \sigma$. This protocol extracts just over $0.5\,\kt$ of work per oscillator relaxation time, consistent with the results of\cc{Archambault-2024}. The trained neural-network protocol \eq{netx} also takes the oscillator position $x$ and the current trap position $x_0$ as input, but is considerably more efficient than its simple counterpart, extracting about 50\% more power. 

The trained neural-network protocol \eq{netv} that takes the oscillator velocity $v$ as input is more efficient still, extracting a mean power of about $0.98\, \kt/\tr$. Significantly, this value almost saturates the bound $\PW \leq \, \kt/\tr$ on extractable power (dotted line) specified by the stochastic thermodynamics of feedback. To derive this bound, consider the heat flow from the bath to the oscillator, which is on average\c{Munakata_2012, Rosinberg_PRE_2017, Dago-2022-JStat}
\beq 
\label{meanQdot}
\mean{\dot \Q} = \frac{\gamma}{m}\kB (T-T_\eff) = \frac{1}{\tr} \kB (T-T_\eff) < \frac{\kt}{\tr},
\eeq
where $T_\eff = m \mean{v^2} / \kB$ is the kinetic temperature of the oscillator. This heat flow is maximum when $T_\eff=0$. In steady state, the extracted power cannot exceed the heat current from the bath, and so the value of \eq{meanQdot} with $T_\eff=0$, namely ${\kt}/{\tr}$, determines the maximum extractable power.

 \f{fig2} illustrates the mechanism underpinning this near-optimal work extraction. Here we show the nature of each protocol (top panels), time traces of the oscillator and trap positions (middle panels), and histograms of extracted work due to each feedback event (bottom panels). 

The neural-network protocol that takes $x$ and $x_0$ as input behaves like a refined version of the simple protocol: the trap position follows the oscillator position, harvesting work in small increments, up to some maximum excursion. For large portions of the trajectory the two coordinates, $x$ and $x_0$, are closely correlated. The probability distribution of work shows that most feedback events extract small amounts of work.

The neural-network protocol that takes $v$ as input behaves differently. The trap position $x_0$ fluctuates strongly, while the oscillator position $x$ is confined near the origin. The work distribution shows a significant number of work-extracting events {\em and} events that require an input of work, and the amounts of work involved are an order of magnitude larger than in the other two protocols. Indeed, the work distribution is almost symmetric, but a slight asymmetry is enough to lead to a net extraction of work that is much larger than for the other protocols. Note that this work-extraction mechanism emerges in the regime of rapid feedback: for increasing feedback time $\tf$, knowledge of the velocity becomes less and less useful, and in the equilibrium (large-$\tf$) limit the extracted power goes to zero: see \f{fig1}(c). This trend contrasts with that using position as an input, which can extract power even in the equilibrium limit\c{Archambault-2025}.

\section{Cold damping}

\begin{figure}
\begin{center}
\includegraphics[width=\linewidth]{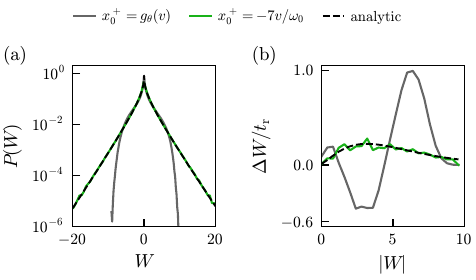}
\end{center}
\caption{(a) Work distribution $P(W)$ for the learned protocol $x_0^+ = g_{\bm \theta}(v)$ (gray) and the linear protocol $x_0^+ = -7v/\omega_0$ (green) that is suggested by the net protocol, compared with the analytic variance-gamma form (black dashed). The analytic prediction (and the numerics derived from the linear protocol) describe a Bessel-function core and exponential tails. (b) Work asymmetry $\Delta W / \tr$, \eqq{asymm}, versus $|W|$. Work $W$ is shown in units of $\kt$.}
\label{fig3}
\end{figure}

Significantly, the neural-network $v$-protocol admits a simple physical description that provides insight into how it extracts power from the thermal bath. To develop this description, note that panel (a, right) of \f{fig2} shows the new trap position $x_0^+$ to be an approximately linear function of cantilever velocity $v$ over the range of trap positions encountered during typical operation (see panel (b, right)). We can therefore approximate the net protocol as
\beq
x_0^+ = -\alpha v,
\eeq
where $\alpha=\alpha_0\omega_0^{-1}$ and $\alpha_0 \approx 7.0$. The feedback time $\tf=2\times10^{-2} \tr$ is short relative to the relaxation time $t_{\rm r}$ of the oscillator, and in order to make progress analytically we assume the limit $\tf \to 0$. In this case we can set $x_0 = -\alpha v$ in \eq{lang} to give
\beq
\label{cd}
m \ddot x + \gamma \dot x + k(x+\alpha \dot x) = \sqrt{2\gamma k_{\rm B}T} \xi(t),
\eeq
which can be rearranged to read
\beq
\label{cd2}
m \ddot x + \gamma_\eff \dot x + kx = \sqrt{2 \gamma_\eff k_{\rm B} T_\eff} \xi(t).
\eeq
\eqq{cd2} is the Langevin equation for a harmonic trap centered at $x=0$, with an effective damping coefficient 
\beq
\gamma_\eff \equiv \gamma+k\alpha=\gamma(1+\alpha_0 Q_f) > \gamma,
\eeq 
and an effective temperature 
\beq
\label{teff}
T_\eff \equiv \frac{\gamma}{\gamma_\eff}T= \left(\frac{1}{1+\alpha_0 Q_f}\right)T < T.
\eeq
These equations describe {\em cold damping}\c{Mancini-1998, Poggio-2007}: an oscillator driven by this particular velocity-dependent feedback force is equivalent to an undriven oscillator with larger damping coefficient and reduced temperature. This strategy is widely used in optomechanics to cool a resonant mode to very low temperature, often for quantum applications\c{Aspelmeyer-2014}. The genetic algorithm has rediscovered this strategy in the course of learning to extract work from the system.

The cold-damped description provides insight into why the demon is able to extract almost the maximum possible amount of power from the thermal bath. Into \eqq{meanQdot} we can insert the effective temperature \eq{teff} to give
\beq
\label{analytic}
\mean{\dot {\Q}}= \frac{\kt}{\tr} \left(1-\frac{1}{1+Q_f\alpha_0}\right)
\approx 0.986 \frac{\kt}{\tr},
\eeq
which is shown as a circle in \f{fig1}(b). The effective temperature is much smaller than the true temperature because the product $Q_f\alpha_0$ is large ($\approx 70$). Thus the heat current from bath to oscillator is close to the theoretical maximum, and so therefore is the extracted power.

The cold-damped description of the system also provides some insight into the fluctuations of extracted work under the net protocol, shown in \f{fig2} panel (c, right). Fluctuations of work extraction under the linear cold-damping protocol $x_0 = -\alpha v$ are described by the {\em variance-gamma} distribution, with a characteristic Bessel-function core and asymmetric exponential tails (\a{appa}). This distribution is sharply peaked near zero and skewed toward positive work, with broad tails that signal large fluctuations. 

In \f{fig3}(a) we show that numerical simulations using the protocol $x_0^+ = -\alpha v$ give rise to a distribution of extracted work values in good agreement with the variance-gamma form. Numerical simulations done using the net protocol \eq{netv} show differences in the tails, which are non-exponential, consistent with the fact that the net protocol is not linear at large values of $|v|$. 

The precise nature of the skew of the work distribution under the neural-net protocol also differs from that of the variance-gamma form. To quantify the asymmetry of the work distribution we consider a set of bins symmetric about the origin. For a bin centered at $|W|$ with edges $W_\pm \equiv |W| \pm \delta W/2$, we define
\beq
\label{asymm}
\Delta W(|W|) \equiv 
\sum_{W_i \in I_-} W_i+\sum_{W_i \in I_+} W_i,
\eeq
where $I_\pm \equiv [ \pm W_\mp, \pm W_\pm]$. We took 25 bins on the interval $|W|<10$, giving $\delta W = 0.4$.

In \f{fig3}(b) we plot $\Delta W/\tr$ versus $|W|$, which shows how different parts of the distribution contribute to the net extracted work. For the linear protocol, the dominant contribution arises at intermediate $|W|$, consistent with the variance-gamma distribution. For the learned protocol the pattern differs, reflecting the protocol's departure from linearity at large $|v|$. 

Thus, the fine details of the neural-network protocol differ from those of the purely linear protocol $x_0^+ = -\alpha v$. However, the peaked center of the distribution and its small skew toward positive work are captured by the analytic expressions obtained by assuming linearity. The linear protocol is also a viable protocol in its own right, extracting essentially the same power as the neural-network protocol. This near-degeneracy suggests that many closely related protocols can achieve near-optimal performance. In this case the value of the neural-network approach is that it revealed the underlying strategy of cold damping.

\section{Energetics and entropy production}

We now examine the energetics and entropy production of the work-extracting velocity feedback, and compare them with those of feedback cooling\c{Kim_PRE_2007,Kim_PRL_2004,Rosinberg_PRE_2017,Munakata_2012}. To simplify the analysis we assume continuous sampling of the feedback.

\subsection{Energetics of cooling}

In previous sections we considered the internal energy $H$ of the system as
\begin{equation}
H = \frac{1}{2}k(x-x_0)^2+\frac{1}{2}mv^2,
\end{equation}
where $x_0$ is the feedback control parameter as defined in \s{Sec_model} for the three kinds of feedback that we tested and compared in the previous section. In this way we consider the force applied by the feedback as an internal force of the system, and this allows us to compare the work of the three controls using the same expression for the work. 
This choice does not present any problem for the first two kinds of feedback where $x_0$ is a nonlinear function of $x$, but can be questioned for the third case where $x_0$ is a function of $v$. In this case we find that 
\begin{eqnarray}
	\dot	\Q & = & -\gamma \dot x^2 + \sqrt{2\gamma k_{\rm B}T} \xi(t) \dot x \\
	 \dot H & = & -\dot W + \dot	\Q\\
	\dot W & = & -\partial_{x_0} U(x,x_0) \ \dot x_0= k (x-x_0) \dot x_0 \nonumber \\
	 &=& k\alpha v^2 -k \alpha \frac{\dd}{\dd t}\left(xv+\frac{1}{2} \alpha v^2\right), \label{Eq_work_our}
\end{eqnarray}
where we used $v=\dot x$	 and the Langevin equation \eq{cd}.

In contrast, in the cooling literature the feedback force $kx_0=-k\alpha v$ is treated as an external nonconservative force, so that the internal energy is
\begin{equation}
H_c =  \frac{1}{2}kx^2+\frac{1}{2}mv^2,
\end{equation}
and one finds that 
\begin{eqnarray}
	\dot	\Q & = & -\gamma \dot x^2 + \sqrt{2\gamma k_{\rm B}T} \xi(t) \ \dot x \\
	\dot	H_c & = & - \dot W_c + \dot \Q \\
	\dot W_c & = & k \alpha v^2 \label{Eq_work}.
\end{eqnarray}

We note that $\dot \Q$ is independent of the convention for the internal energy, whereas the extracted power $\dot W$ and $\dot W_c$ differ by a total derivative term. The mean power $\av{\dot W}=\av{\dot W_c}$ is thus independent of the convention, but its fluctuations are not. In particular, $\dot W_c=k\alpha v^2$ is always positive, whereas $\dot W$ exhibits large positive and negative fluctuations (see \f{fig2}).

\subsection{Entropy production}

The fact that the heat is independent of whether we use $H$ or $H_c$ has important consequences for the entropy production rate in feedback cooling. This problem has been discussed in several references\c{Kim_PRE_2007,Kim_PRL_2004}. The total entropy production rate $\dot S_p$ is
\begin{eqnarray} \label{Eq_entropy_rate}
	\dot S_p &=& -{ \dot \Q \over T} - \frac{\dd}{\dd t} \ln\left( P(x,v) \right)-\dot S_{pu},
\end{eqnarray}
where $S_{pu}$ is the pumping entropy\c{Kim_PRL_2004}, present only when the feedback depends on $v$: $\dot S_{pu}= \kB \omega_0^2 \partial_v g(v)$ with $g(v)=-\alpha v$ or $g(v)=g_{\bm \theta}(v)$ in our case. The total entropy production in a time interval $\tau$ is $\Delta S_p(\tau)=\int_0^\tau \dot S_p \dd t=S_p(\tau)-S_p(0)$ which according to\ccc{Kim_PRE_2007, Munakata_2012, Rosinberg_PRE_2017} satisfies an integral fluctuation theorem (IFT) i.e.:
\begin{eqnarray} \label{Eq_Int_S}
	\Mean{\exp\left(-\frac{\Delta S_p(\tau)}{k_{\rm B}}\right)}=1
\end{eqnarray}
which imposes $\av{\Delta S_p(\tau)}\ge 0$. Note that $S_{pu}$ is essential for this inequality and \eqq{Eq_Int_S} to hold. 
The other FT concerns $\Q_\tau=\int_0^\tau \dot \Q \dd t$: 
\begin{eqnarray} \label{Eq_Int_Q}
	\Mean{\exp\left(\frac{\Q_\tau}{k_{\rm B}T}-\frac{\gamma}{m} \tau\right)}=1
\end{eqnarray}

This equation imposes not only $\av{\Q_\tau/\tau}\le \gamma k_{\rm B}T /m $ which is equivalent to \eqq{meanQdot}, but also strong constraints on the fluctuations of $\Q_\tau$. We checked  to what extent  Eqs.~\eq{Eq_entropy_rate} and~\eq{Eq_Int_Q} are satisfied when a finite $\tf$ is used instead of a continuous sampling (CS). From here on we present results with the cold-damping scheme, i.e. $g(v)=-\alpha v$, but equivalent observations were obtained with a hyperbolic-tangent approximation of $g_{\bm \theta}(v)$. We computed the FT at various $\tau$ in the range $[\tf,2\pi/\omega_0]$. The value of the FT is estimated from the mean over an ensemble of 50 subsets of $2\times10^4$ trajectories of length $\tau$. The error bar is estimated from the standard deviation of the values of the IFT computed on each subset.  In \f{fig4}(a) we plot the results of the IFT for $\Q_\tau$ at $\tf=2\times10^{-2}\tr$ and $\alpha_0=7$. We clearly see that \eqq{Eq_Int_Q} is satisfied even for a finite $\tf\ll \tr$. 
\begin{figure}
	\centering
	\includegraphics[width=\linewidth]{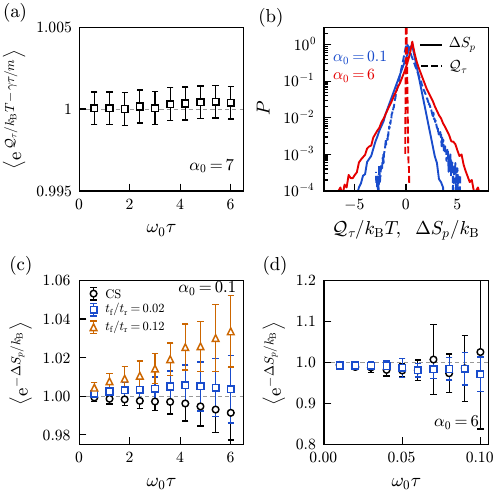}
	\caption{Fluctuation theorems for the cold-damped dynamics, with dimensionless time $\omega_0\tau$ on the horizontal axes. (a)~Integral fluctuation theorem (IFT) for the heat $\Q_\tau$ at $\alpha_0=7$ and $\tf=2\times10^{-2}\tr$. (b)~Distributions $P(\Q_\tau/\kt)$ (dashed) and $P(\Delta S_p/\kB)$ (solid) at $\alpha_0=0.1$, $\omega_0\tau=6$ (blue) and $\alpha_0=6$, $\omega_0\tau=0.1$ (red), chosen so that the pumping entropy $S_{pu}$ is equal in the two cases. (c)~IFT for the entropy production $\Delta S_p$ at $\alpha_0=0.1$, for continuous sampling (CS) and two finite feedback times $\tf$. (d)~Same as (c) at $\alpha_0=6$, for CS and one finite $\tf$; the fluctuations are so large that no conclusion about the role of $\tf$ can be drawn.}
	\label{fig4}
\end{figure}

Studying the effect of a finite $\tf$ on the IFT for the entropy is more difficult, because the entropy fluctuations increase with $\alpha_0$. To illustrate this problem we plot in \f{fig4}(b)  the pdf of $\Delta S_p(\tau)$ and $\Q_\tau$ at  $\alpha_0=0.1$ and $\alpha_0=6$ corresponding to $T_\eff=0.5\,T$  and $T_\eff\approx0.016\,T$ respectively. For $\tau$ we take values such that $S_{pu}$  is the same for the two values of $\alpha_0$, thus $\tau=6$ at $\alpha_0=0.1$ and  $\tau=0.1$ for $\alpha_0=6$. Although $S_{pu}$ is the same in the two cases, the fluctuations have very different statistics. At $\alpha_0=0.1$, $\Q_\tau$ has very small fluctuations comparable to those of the entropy. By contrast, at $\alpha_0=6$ the fluctuations of $\Q_\tau$ are very small whereas those of the entropy increase strongly because of the small $T_\eff$, i.e. the entropy fluctuations are dominated by those of the system entropy $-\ln[P(x(\tau),v(\tau))/P(x(0),v(0))]$. Therefore the IFT fluctuations are very large, the mean being dominated by the extreme negative fluctuations of $\Delta S_p(\tau)$. In \f{fig4}(c,d) we plot the computed  IFT for $\Delta S_p(\tau)$ as a function of $\tau$ for various $\tf$. In panel (c) where $\alpha_0=0.1$, we observe a drift as a function of $\tau$, but a very small error in the measured value of $T_\eff$ can account for it. The situation is worse for $\alpha_0=6$
as can be seen in panel (d). The fluctuations are so large that it is impossible to draw any conclusion about the role of $\tf$ in the IFT.  

\section{Conclusions}

We trained a neural-network Maxwell's demon to extract work from a model of an underdamped micromechanical cantilever subject to thermal noise. The demon, which periodically adjusts the position of a harmonic trap, is trained to maximize the power extracted under steady-state operation. When the demon is given the cantilever position and trap position as inputs it learns a refined version of an existing hand-designed protocol, yielding a substantial improvement in performance. When the demon receives the oscillator velocity as input it discovers a qualitatively different strategy that extracts substantially more work, close to the theoretical power bound. 

Analysis of the protocol shows that it implements {\em cold damping}: the trap position is displaced approximately linearly with velocity, producing an effective increase of the oscillator's damping coefficient and a reduction of its effective temperature. This strategy, well known from the field of optomechanics, reveals a simple physical mechanism underlying near-optimal work extraction from thermal fluctuations in an underdamped system. 

\section{Acknowledgments} SW performed work as part of a user project at the Molecular Foundry at Lawrence Berkeley National Laboratory, supported by the Office of Basic Energy Sciences of the U.S. Department of Energy under Contract No. DE-AC02--05CH11231. SW was partially supported by the US Department of Energy, Office of Science, Office of Basic Energy Sciences Data, Artificial Intelligence and Machine Learning at DOE Scientific User Facilities program under Award Number 34532 (a digital twin for in silico spatiotemporally-resolved experiments). LB acknowledges funding provided by project ANR-22-CE42-0022. 

\appendix

\section{Distribution of instantaneous power} 
\label{appa}

Analysis of the effective description of the neural-network $v$-protocol, \eqq{cd2}, provides some insight into the shape of the distribution shown in \f{fig2}(c, right). At the outset we note that \eqq{cd2} assumes infinitely rapid feedback, whereas in practice the trap position is changed at finite-time increments. We therefore assume an implicit regularization time $\Delta$ in the subsequent analysis. 

Writing $v = \dot{x}$ we have
\bea
\label{cda}
\dot x &=& v, \\
\label{cdb}
m\dot v &=& -\gamma_{\rm eff}\,v - kx + \sqrt{2\gamma_{\rm eff} \kB T_{\rm eff}}\,\xi(t),
\eea
where $\xi(t)$ is a Gaussian white noise with correlations $\av{\xi(t)} = 0$ and $\av{\xi(t) \xi(t')} = \delta(t-t')$.

Because these equations are linear in $x$ and $v$ and driven by additive Gaussian white noise, the steady-state dynamics is an Ornstein-Uhlenbeck process, and the joint distribution of $(x,v)$ is Gaussian. Specifically,
\beq
x \sim \mathcal{N}\bigl(0,\langle x^2\rangle\bigr), \quad
v \sim \mathcal{N}\bigl(0,\langle v^2\rangle\bigr),
\eeq
with steady-state moments
\beq
\label{moments}
\langle x^2\rangle = \frac{\kB T_{\rm eff}}{k} , \quad \langle v^2\rangle = \frac{\kB T_{\rm eff}}{m},
\eeq
and 
\beq
 \langle x v\rangle=\frac{1}{2} \frac{{\rm d}}{{\rm d}t}\av{x^2}=0.
\eeq 
\eqq{moments} shows that cold damping suppresses fluctuations of position and velocity relative to the system in the absence of feedback (recall that $T_{\rm eff}<T$).

It is convenient to define the coordinate
\beq
y \equiv x-x_0 = x+\alpha v,
\eeq
in terms of which the potential energy is 
\beq
U(x,x_0)=\frac{1}{2} k y^2.
\eeq
The instantaneous power delivered by the trap is then
\beq
\label{power}
\PW(t) = -\partial_{x_0} U(x,x_0) \dot x_0 = -k\alpha\, y(t)\dot v(t).
\eeq
If $(y,\dot v)$ were zero-mean correlated Gaussian variables then their product would follow a {\em variance-gamma} (VG) distribution\c{fischer2025variance}, such that
\bea
\label{prob1}
f(\PW)&=&
\frac{1}{\pi\,k\alpha\,\sigma_y\,\sigma_{\dot v}\sqrt{1-\rho^2}}\,
\exp\!\Big[\frac{\rho\,\PW}{k\alpha(1-\rho^2)\sigma_y\sigma_{\dot v}}\Big] \nonumber\\
&\times&
K_0\!\Big(\frac{|\PW|}{k\alpha(1-\rho^2)\sigma_y\sigma_{\dot v}}\Big).
\label{vg-yvdot}
\eea
Here $K_0$ is the modified Bessel function of the second kind, and the correlation coefficient $\rho$ is
\beq
\label{rho}
\rho = -\frac{\av{y\dot v}}{\sigma_y\sigma_{\dot v}},
\eeq
which is defined in terms of the moments
\bea
\sigma_y^2 &\equiv& \langle y^2\rangle
 = \left(\frac{1}{k}+\frac{\alpha^2}{m}\right) \kB T_{\rm eff}, \\
\sigma_{\dot v}^2 &\equiv& \langle \dot v^2\rangle, \quad {\rm and} \label{middle}\\
\langle y\,\dot v\rangle &=&- \frac{\kB T_{\rm eff}}{m}. \label{Eq.mean.yvdot}
\eea
However, while $y(t)$ is an ordinary Gaussian random variable, $\dot v(t)$ is a generalized stochastic process that depends on delta-correlated noise. Consequently, $\PW(t)$ should be understood as a distribution-valued quantity requiring short-time regularization. Such regularization is also physically appropriate, because in simulations we extract work in discrete increments rather than continuously. We therefore introduce a finite time increment $\Delta$ to define a regularized power. While the feedback time $\tf$ provides a natural candidate for $\Delta$, it does not give the best quantitative agreement with numerics. In practice, we treat $\Delta$ as a phenomenological short-time cutoff.

We must regularize $\sigma_{\dot v}^2=\av{\dot v^2}$, which we choose to do by defining the regularized acceleration
\beq
\dot v \to a_\Delta(t) \equiv \frac{v(t+\Delta)-v(t)}{\Delta},
\eeq
whose variance 
\beq
\sigma_{a_\Delta}^2 \equiv \langle a_\Delta^2\rangle
 = \frac{2}{\Delta^2}\bigl[\langle v^2\rangle - C_v(\Delta)\bigr]
\eeq
is finite and determined by the steady-state mean-square velocity $\av{v^2}= \kB T_{\rm eff}/m$ and the velocity autocovariance 
\beq
\label{cv}
C_v(t) \equiv \langle v(t)v(0)\rangle = \frac{\kB T_{\rm eff}}{m} g_v(t).
\eeq
Here
\bea
g_v(t) &\equiv& \frac{a_+\e^{-a_+\omega_0 t} - a_-\e^{-a_-\omega_0 t}}{a_+-a_-}, \, \, {\rm and} \label{eq.gv}\\
a_\pm & \equiv& \frac{(\alpha_0+1/Q_f)\pm\sqrt{(\alpha_0+1/Q_f)^2-4}}{2}.\label{eq.apm}
\eea
Since $\alpha_0\approx7\gg 1/Q_f=0.1$, we have $a_+\approx \alpha_0$ and $a_-\approx 1/\alpha_0$. The derivation of Eqs.~\eq{cv}--\eq{eq.apm} is given in \a{appb}. 

Thus the correlation coefficient $\rho$, \eqq{rho}, is
\beq
\rho=\frac{\Delta \omega_0}{\sqrt{2 \left(1+\alpha_0^2 \right) (1-g_v(\Delta))}}.
\eeq
Note that for small $\Delta$, $\sigma_{a_\Delta}^2$ diverges as $1/\Delta$ and $\rho$ vanishes as $\sqrt \Delta$:
\beq
\rho \approx \sqrt{\frac{\Delta \omega_0}{2 \left(1+\alpha_0^2 \right)\left(\alpha_0+1/Q_f\right)}}.
\eeq

With these details in hand, we can derive insight into several of the features of the work distribution shown in \f{fig2}(c, right). 

First, the mean instantaneous power extracted by the feedback follows from \eq{power}, and is
\bea
\langle \PW \rangle
 = -k\alpha\,\langle y\,\dot v\rangle
 &=& \alpha k
 \frac{\kB T_{\rm eff}}{m} \\[4pt]
 &=& \frac{\gamma}{m}\,k_{\rm B}\bigl(T - T_{\mathrm{eff}}\bigr),
\eea
using $T_\eff \equiv T \gamma/\gamma_\eff$. This result agrees with \eqq{analytic} (the minus sign is introduced to conform to the sign convention of the main text).

Second, the approximate shape of the numerical work distribution can be understood from the factors of \eqq{prob1}. The modified Bessel function $K_0$ produces the sharp peak near $\PW=0$. The asymmetry of the distribution is controlled by the exponential prefactor and the correlation coefficient $\rho$. The distribution is skewed toward positive $\PW$, corresponding to net energy extraction: the feedback extracts more work from the bath than it injects. The analytic distribution $P(W)$ shown in \f{fig3} is obtained from $f(\PW)$ via the relation $P(W) = t_{\rm f}^{-1} f(W/t_{\rm f})$.

Quantitatively, the agreement with numerics is not exact: the tails and the precise degree of asymmetry are not captured in detail. This discrepancy arises because the learned protocol is not strictly linear in $v$: it oscillates around a linear form and saturates for large $|v|$, whereas the analytic theory assumes $x_0=-\alpha v$. However, when we replace the learned protocol by this linear feedback protocol in simulations, the resulting work distribution is indeed well described by the variance-gamma form (\f{fig3}). Thus the analytic theory captures the essential mechanism and the main features of the work fluctuations within the ideal cold-damped description. The version of cold damping learned by the neural-net demon differs enough that its fine details differ from those of the linear protocol, but several features of the neural-net protocol can nonetheless be understood using the analytic theory.

\section{Velocity autocovariance}
\label{appb}

To compute \eqq{cv}, $C_v(t) = \av{v(t) v(0)}$, differentiate it with respect to time to give
\beq
\dot C_v(t)=\frac{d}{dt}\langle v(t)v(0)\rangle
=\langle \dot v(t)\,v(0)\rangle.
\label{Cv1}
\eeq
Using \eq{cdb} and the fact that $\xi(t>0)$ and $v(0)$ are uncorrelated gives
\beq
\dot C_v(t)
= -\Gamma C_v(t)-\omega_0^2\,C_{xv}(t),
\label{Cv2}
\eeq
where $\Gamma\equiv\gamma_{\eff}/m=(\alpha_0+1/Q_f)\omega_0$ and $C_{xv}(t)\equiv \langle x(t)v(0)\rangle$. Next, differentiate $C_{xv}(t)$ to give
\beq
\dot C_{xv}(t)=\frac{d}{dt}\langle x(t)v(0)\rangle
=\langle v(t)v(0)\rangle
=C_v(t).
\label{Cxv1}
\eeq
Differentiating \eq{Cv2} with respect to time and eliminating $\dot{C}_{xv}(t)$ using
\eq{Cxv1} gives a closed equation for $C_v(t)$, namely
\beq
\ddot C_v(t)+\Gamma\,\dot C_v(t)+\omega_0^2\,C_v(t)=0,
\label{CvODE}
\eeq
whose initial conditions are $C_v(0)=\langle v^2\rangle=\kB T_{\rm eff}/m,$ and
\beq
\dot C_v(0)=\langle \dot v(0)\,v(0)\rangle
= -\Gamma\langle v^2\rangle-\omega_0^2\langle x v\rangle
= -\Gamma\langle v^2\rangle.
\label{CvIC2}
\eeq
Solving \eq{CvODE} in the overdamped regime $\omega_0<\Gamma/2$ gives \eqq{cv}.


%

\end{document}